\documentclass[11pt]{cernrep}
\usepackage{graphicx}
\usepackage{amsmath}
\newcommand{\unity}{1\hspace{-1.3mm}1}

\begin{document}

\title{Phase diagrams of the three-flavor NJL model with
color superconductivity and pseudoscalar condensation}

\author{Harmen J. Warringa}

\institute{Department of Physics and Astronomy, Vrije Universiteit, \\
De Boelelaan 1081, 1081 HV Amsterdam, The Netherlands}

\maketitle

\begin{abstract}
We present results of the calculation of phase diagrams of the
three-flavor NJL model as a function of temperature and different
quark chemical potentials. These phase diagrams are an extension of
earlier calculations in the literature in which either only the
possibility of color superconductivity or the possibility of
pseudoscalar condensation was investigated. This study takes
into account both options and hence allows one to investigate their
competition. It turns out that the color superconducting phase and the
phase in which the pseudoscalar mesons condense are separated by a
first order transition.
\end{abstract}

\section{Introduction}
The NJL model is a low-energy effective theory for QCD and as such
widely used for investigating the phase structure of QCD as a function
of temperature and chemical potentials. Many studies have been carried
out for equal quark chemical potentials, that is $\mu_u = \mu_d =
\mu_s$. This constraint does not necessarily hold in realistic
situations. For example, in a compact star in equilibrium under the
weak interactions $\mu_d = \mu_s$, and $\mu_u$ is tuned in such a way
that the star becomes electrically neutral \cite{Alford2002,
Ruster2005, Blaschke2005}. Therefore, it is useful and interesting to
investigate the phase diagram of the NJL model as a function of
different chemical potentials.  It turns out that the NJL model has a
rich phase structure with phases in which chiral symmetry is broken,
different color superconducting phases and phases in which the
pseudoscalar mesons condense. In the last mentioned phase a field with
the same quantum numbers as the pion or the kaon obtains a vacuum
expectation value, that is the condensates $\langle \bar u i \gamma_5
d \rangle$, $\langle \bar u i \gamma_5 s \rangle$ or $\langle \bar d i
\gamma_5 s \rangle$ are nonzero. The interesting feature of these
phases is that parity is broken there. The existence of a pion
condensate \cite{Son2001} is also confirmed on the lattice at finite
isospin chemical potential with zero baryon chemical potential
\cite{Kogut2002}.  The analysis presented in this article (and in more
detail in Ref.\ \cite{Warringa2005}) is an extension of previous
studies in which either only pseudoscalar condensation
\cite{Toublan2003, Barducci2004, Barducci2005} or only color
superconductivity \cite{Gastineau2002, Neumann2003, Lawley2005} was
taken into account. Here both possibilities will be investigated
simultaneously. The more complete phase diagrams which we have
calculated are not merely a superposition of the phase diagrams
obtained earlier, since it turns out that there is competition between
the color superconducting phases and the phases in which the
pseudoscalar mesons condense.  If the chemical potentials of the
quarks are large and approximately equal, for example $\mu_u \approx
\mu_d$, it is possible that an up and a down quark form a diquark
condensate giving rise to a two-flavor color superconducting (2SC)
phase.  If the chemical potentials of the quarks are large and
approximately opposite to each other, for example $\mu_u \approx
-\mu_s$ it is possible that an up and a strange quark form a charged
kaon condensate $\langle \bar u i \gamma_5 s \rangle$. So one might
wonder what would happen if $\mu_u \approx \mu_d \approx -\mu_s$,
could a 2SC phase appear together with a phase in which the charged
kaon meson condenses? We find that the color superconducting phases
and the phase in which the pseudoscalar mesons condense are always
separated by a first order transition for any temperature for a set of
realistic parameters chosen.

The three-flavor NJL model we use is derived from a 4-point color
current-current interaction and given by the following Lagrangian
density (see for example Ref.\ \cite{Buballa2005a})
\begin{multline}
  \mathcal{L} = \bar \psi \left(i \gamma^\mu \partial_\mu - M_0
  + \mu \gamma_0 \right)\psi 
  + G \left(\bar \psi \lambda_a \psi \right)^2 + 
 G \left(\bar \psi \lambda_a i \gamma_5 \psi \right)^2 
\\ 
 + \frac{3}{4} G \left(\bar \psi t_{A} \lambda_{B} C i \gamma_5 \bar \psi^T \right)
\left(\psi^T  t_{A} \lambda_{B} C i \gamma_5 \psi \right)
 \label{eq:lagrnjl} \;,
\end{multline}
where $G$ is a coupling constant, $M_0$ is a diagonal matrix
containing the up, down and strange quark masses and $\mu$ a diagonal
matrix containing the up, down and strange quark chemical
potentials. The constants $A, B \in \{2, 5, 7\}$, since the one gluon
exchange interaction is only attractive in the antisymmetric color and
flavor triplet channel.  The matrices $\lambda_a$ are the $9$
generators of $\mathrm{U}(3)$ and act in flavor space. They are
normalized as $\mathrm{Tr}\, \lambda_a \lambda_b = 2 \delta_{a b}$.
The matrices $t_a$ are the generators of $\mathrm{U}(3)$ and act in
color space.  Their normalization is $\mathrm{Tr}\, t_a t_b = 2
\delta_{a b}$.  To remind the reader, the antisymmetric flavor
matrices $\lambda_2$, $\lambda_5$ and $\lambda_7$ couple up to down,
up to strange and down to strange quarks, respectively.  The charge
conjugate of a field $\psi$ is denoted by $\psi_c = C \bar \psi^T$
where $C = i \gamma_0 \gamma_2$.  The coupling strength $3 G /4$ of
the diquark interaction is fixed by the Fierz transformation (see for
example Ref.\ \cite{Buballa2005a}). However, other choices for this
coupling strength are also made in the literature.

The NJL model has a symmetry structure similar to QCD, except for the
local color symmetry. The NJL model is only invariant under global SU(3)
color transformations since it does not contain gauge fields.  In
absence of quark masses and chemical potentials, the Lagrangian
density of the NJL model has a global
$\mathrm{SU(3)}_c\times\mathrm{U(3)}_{L}\times\mathrm{U(3)}_{R}$
symmetry. Due to the non-vanishing quark masses, the symmetry is like
in QCD broken down to $\mathrm{SU(3)}_c \times
\mathrm{U(3)}_{V}$. Since the masses of the quark and the chemical
potentials are chosen to be different, the symmetry of the Lagrangian
density is further reduced to
$\mathrm{SU(3)}_c\times\mathrm{U(1)}_B\times\mathrm{U(1)}_I
\times\mathrm{U(1)}_Y $, where $B$, $I$, $Y$ stand respectively for
baryon, isospin (the $z$-component) and hypercharge number. These
remaining symmetries can be used to rotate away several possible
condensates which simplifies the analysis in the end.

The results that will be presented here are obtained with the
following choice of parameters which were also used in Ref.\ 
\cite{Buballa2005a}
\begin{equation}
\begin{array}{cccc}
 m_{0u} = m_{0d} =  5.5\;\mathrm{MeV}, & m_{0s} =
112\;\mathrm{MeV}, &
 G = 2.319 / \Lambda^2, & \Lambda = 602.3\;\mathrm{MeV}.
\end{array}
\end{equation}
These are the precise values used in the calculations, but clearly
not all digits are significant implying that a small change
of parameters will not have a big influence on the results.
This choice of parameters gives rise to constituent quark masses
$M_u=M_d=368$ MeV and $M_s=550$ MeV. 

\section{Effective potential}
To obtain a phase diagram one has to investigate the behavior of order
parameters like the chiral condensates $\alpha_a =
-2 G \langle \bar \psi \lambda_a \psi \rangle$, the pseudoscalar
condensates $\beta_a = -2 G \langle \bar \psi i \gamma_5 \lambda_a
\psi \rangle$ and the diquark condensates $\Delta_{AB} = \frac{3}{2}G
\langle \psi^T t_A \lambda_B C \gamma_5 \psi \rangle$. The values of
these order parameters can be found by minimizing the effective
potential. The effective potential can be obtained by the introduction
of auxiliary fields which are then shifted in such a way that the
action of the NJL becomes quadratic in the fermion fields.  After
integration over the fermion fields and replacing the auxiliary fields
by their vacuum expectation values, it turns out \cite{Warringa2005}
that the effective potential in the mean field approximation is
given by
\begin{multline}
  \mathcal{V} = \frac{\alpha_a^2 + \beta_a^2}{4G}
+ \frac{\left \vert \Delta_{AB}\right \vert^2 }{3 G} 
\\ -
\frac{T}{2} 
\sum_{p_0 = (2n + 1)\pi T} 
\int \frac{\mathrm{d}^3 p}{
\left( 2\pi \right)^3} \log \mathrm{det} 
\left(
\begin{array}{cc}
\unity_c \otimes \mathcal{D}_1 
&
\Delta_{AB}\, t_A \otimes \lambda_B \otimes \gamma_5
\\
-\Delta^*_{AB}\,  t_A \otimes \lambda_B \otimes \gamma_5 
&
\unity_c \otimes \mathcal{D}_2
\end{array} \right),
\label{eq:njleffpotential}
\end{multline}
where
\begin{eqnarray}
  \!\!\!\! \mathcal{D}_1 \!\!\!\!&=& \!\!\!\!\unity_f \otimes (i \gamma_0 p_0\!+\!
\gamma_i p_i) 
-  \mu \otimes \gamma_0 
  - ( M_0\!+\!\alpha_a \lambda_a ) \otimes \unity_d 
- \beta_a \lambda_a \otimes i \gamma_5 ,
\\
  \!\!\!\! \mathcal{D}_2 \!\!\!\! &=& \!\!\!\! \unity_f \otimes (i \gamma_0 p_0\!+\!\gamma_i
p_i) 
 +  \mu \otimes \gamma_0
 - ( M_0\!+\!\alpha_a \lambda_a^T)\otimes \unity_d  
 - \beta_a \lambda_a^T \otimes i \gamma_5
.
\end{eqnarray}
The matrix $\unity$ is the identity matrix in color ($c$), flavor ($f$),
or Dirac ($d$) space.

By minimizing the effective potential it was found numerically that of
the chiral condensates only $\alpha_0$, $\alpha_3$ and $\alpha_8$ can
be nonzero. So the chiral condensates $\langle \bar u u \rangle$,
$\langle \bar d d \rangle$ and $\langle \bar s s \rangle$ appear,
while for example a $\langle \bar u d \rangle$ condensate is
impossible.  The chiral condensates can be found in the phase diagram
at low chemical potentials and temperatures.

Furthermore, it turns out that the pseudoscalar condensates $\beta_0$,
$\beta_3$ and $\beta_8$ are always nonzero. However, the other
pseudoscalar condensates can appear. Using the $\mathrm{U}(1)$ flavor
symmetries one can reduce the possible set of pseudoscalar condensates to
$\beta_{2}$, $\beta_5$ and $\beta_7$, which are respectively a pion
($\pi^\pm$), a charged kaon ($K^\pm$), and a neutral kaon $K^0 / \bar
K^0$ condensate. All these condensates break parity. Pion condensation
is possible if $\vert \mu_u - \mu_d \vert > m_\pi$ \cite{Son2001}
where $m_\pi$ is the mass of the pion, while kaon condensation is
possible if $\vert \mu_{u,d} - \mu_s \vert > m_K$ \cite{Kogut2001}
where $m_K$ is the kaon mass.

The different possible color-superconducting phases are
named as follows (see for example \cite{Ruster2005})
\begin{equation}
\begin{array}{ccclcccl}
 \Delta_{22} \neq 0, &\Delta_{55}\neq 0, &\Delta_{77} \neq 0 & \mathrm{CFL} \;,
 \hspace{1.0cm} &
 \Delta_{22} \neq 0, & \Delta_{55}=0, & \Delta_{77} = 0 & \mathrm{2SC} \;,
\\
 \Delta_{77} = 0, & \Delta_{22} \neq 0, & \Delta_{55} \neq 0  & \mathrm{uSC} \;,
&
 \Delta_{55} \neq 0, & \Delta_{22} = 0, & \Delta_{77} = 0 & \mathrm{2SCus} \;,
\\
 \Delta_{55} = 0, & \Delta_{22}\neq0, & \Delta_{77} \neq 0  & \mathrm{dSC} \;,
 &
 \Delta_{77} \neq 0, & \Delta_{22}=0, & \Delta_{55} = 0 & \mathrm{2SCds} \;.
\\
 \Delta_{22} = 0, & \Delta_{55} \neq 0, & \Delta_{77} \neq 0  & \mathrm{sSC} \;.
 & & & & 
\end{array}
\end{equation}
The abbreviation CFL stands for color-flavor locked phase. If there is
exact $\mathrm{SU}(3)_\mathrm{V}$ flavor symmetry, the three diquark
condensates in this phase have equal size and the vacuum is invariant
under a combined rotation in color and flavor space \cite{Alford1999}.
In the uSC, dSC or sSC phase the up, down or strange quark always
takes part in the diquark condensate, respectively. In the 2SC phase
an up and a down quark form a diquark condensate, in the 2SCus and the
2SCds phase this condensate is formed by an up and and strange quark
and a down and a strange quark, respectively.  The color
superconducting phases of the NJL model are not color neutral. In
order to achieve color neutrality, one should introduce color
chemical potentials \cite{Steiner2002, Buballa2005c}. However, we
leave this for further work.

To calculate the effective potential one needs to evaluate a
determinant of a $72 \times 72$ matrix. Only in special cases such as
when all masses and chemical potentials are equal and in absence of
pseudoscalar condensation one can perform the sum over Matsubara
frequencies analytically and hence simplify the effective potential
somewhat further. But in the more general cases which will be
discussed in this article this either is not possible or gives rise to
very complicated equations. We therefore calculated and minimized the
effective potential numerically.

\section{Phase diagrams}
\begin{figure}[t]
\begin{tabular}{cc}
\scalebox{0.9}{\includegraphics{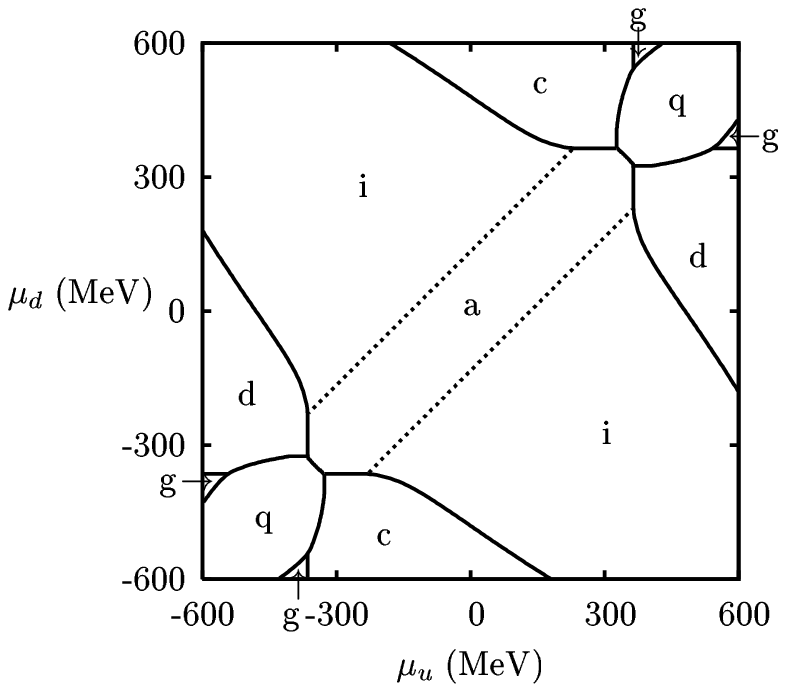}} &
\scalebox{0.9}{\includegraphics{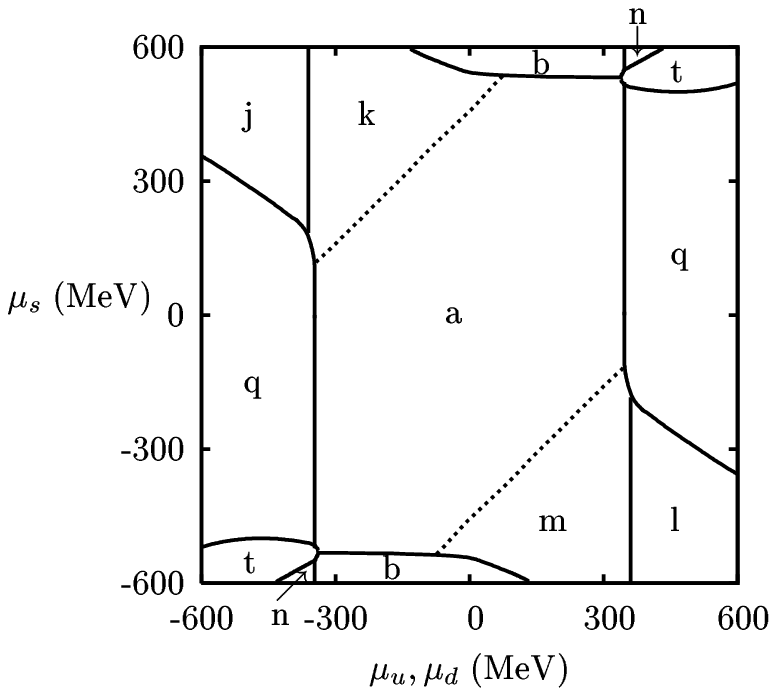}}\\
\scalebox{0.9}{\includegraphics{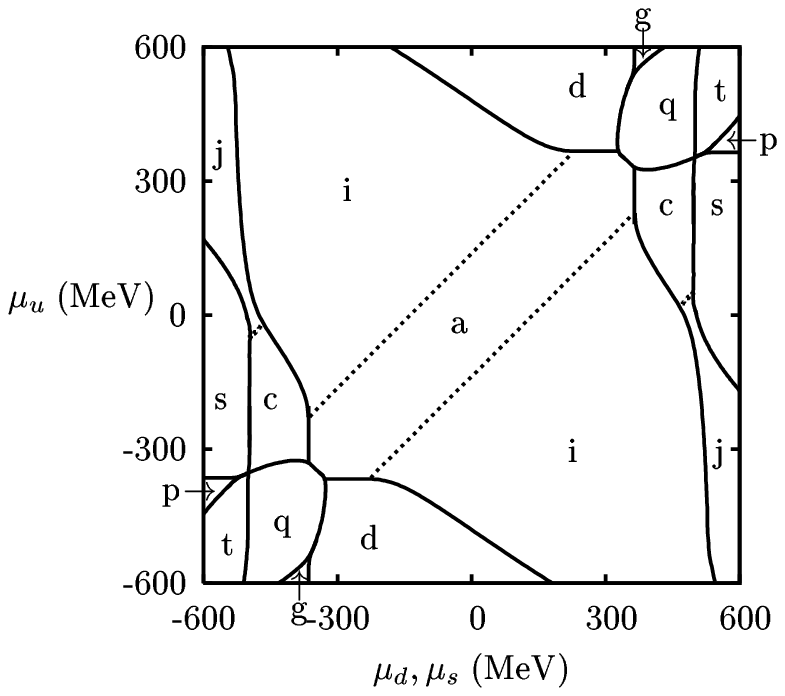}} &
\scalebox{0.9}{\includegraphics{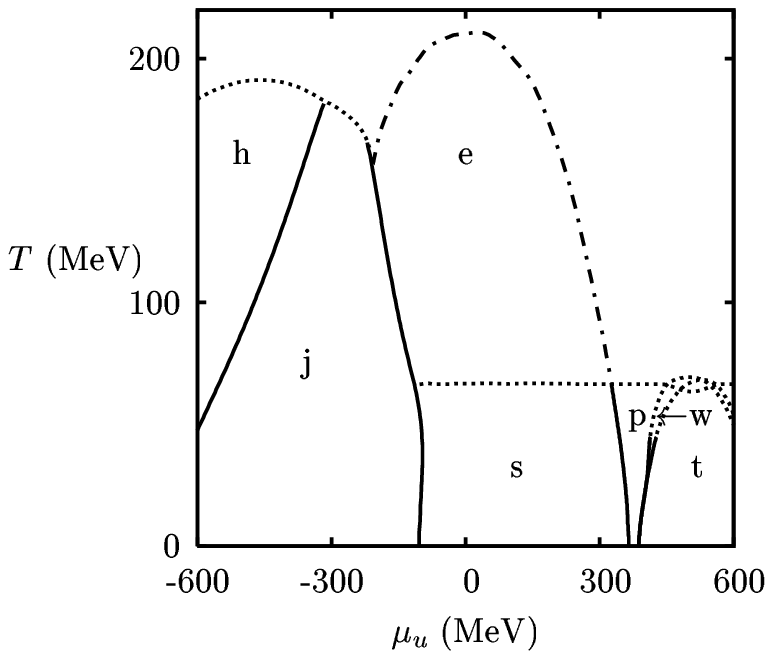}}
\end{tabular}
\caption{Phase diagrams of the NJL model for $\mu_s = 0$ and $T=0$
(upper left), $\mu_u=\mu_d+\epsilon$ and $T=0$ (upper right), $\mu_d = \mu_s$
and $T=0$ (lower left), and $\mu_d = \mu_s = 550\;\mathrm{MeV}$ (lower
right).  First and second-order transitions are indicated by solid and
dotted lines, respectively, while cross-overs are denoted by
dashed-dotted lines. The letters denote the different phases, where 
a: $\bar u u$ + $\bar d d$ + $\bar s s$, 
b: $\bar u u$ + $\bar d d$, 
c: $\bar u u$ + $\bar s s$, 
d: $\bar d d$ + $\bar s s$,
e: $\bar u u$,
g: $\bar s s$, 
h: $\pi^+ / \pi^-$, 
i: $\pi^+ / \pi^-$ + $\bar s s$, 
j: $K^+ / K^-$, 
k: $K^+/ K^-$ + $\bar d d$,
l: $K^0 / \bar K^0$, 
m: $K^0/\bar K^0$ + $\bar u u$,
n: 2SC,
p: 2SCds,
q: 2SC + $\bar s s$,
r: 2SCus + $\bar d d$, 
s: 2SCds + $\bar u u$, 
t: CFL and
w: sSC.}
\label{fig:diagrams}
\end{figure}

Some of the phase diagrams we obtained in Ref.\ \cite{Warringa2005}
are displayed in Fig.\ \ref{fig:diagrams}. Our results agree
qualitatively with the phase diagrams with only pseudoscalar
condensation presented in Refs.\ \cite{Barducci2004,Barducci2005} and
the phase diagram with only superconductivity presented in Ref.\
\cite{Gastineau2002}.

One can clearly see that the phase diagrams evaluated at $T=0$ are
symmetric under reflection in the origin. This is because the free
energy is invariant under the transformation $(\mu_u, \mu_d, \mu_s)
\rightarrow (-\mu_u, -\mu_d, -\mu_s)$ from the symmetry between
particles and anti-particles. 

In general, horizontal and vertical lines in the phase diagrams arise
if the pairing of one type of quark is not changed after a
transition. In this case, the location of the phase boundary is
determined by the properties of other quarks. Therefore, changing the
chemical potential of the unchanged quark species cannot have a big
influence on the location of the phase boundary. This results in the
horizontal and vertical lines. For $T=0$, one always finds these lines
near the values of the constituent quark masses, i.e.\ $\mu_u \approx
M_u$, $\mu_d \approx M_d$ and $\mu_s \approx M_s$ (see for example
Ref.\ \cite{Buballa2005a}). The diagonal lines arise because at $T=0$
pion condensation can occur if $\vert \mu_u - \mu_d \vert > m_\pi =
138\;\mathrm{MeV}$ \cite{Son2001} and kaon condensation can occur if
$\vert \mu_s - \mu_{u, d} \vert > m_K = 450\;\mathrm{MeV}$
\cite{Kogut2001} (the chosen parameter set gives rise to a somewhat
low kaon mass, but this is not relevant for the qualitative features
of the phase diagram).

The diagrams show that if the chemical potentials are different from
each other, the transition to the color-supercon\-ducting phases
(n/p/q/t/s) remains first order at $T=0$ as was concluded in Ref.\
\cite{Bedaque2002}.  The lower right figure shows that this conclusion
does no longer hold at finite temperature.

Moreover, one can see from the upper left diagram that if $\mu_u \neq
\mu_d$ it is possible to go through two first-order transitions before
entering the 2SC phase (q) (similar to the situation discussed in
Ref.\ \cite{Toublan2003} without color superconductivity). We observe
that to have such a scenario at zero temperature, a minimum difference
between $\mu_u$ and $\mu_d$ is required. In the present case this is
about 35 MeV.  Pion condensation (i) and the 2SC phase (q) are in this
diagram separated by two phase transitions in contrast to the
estimated $(\mu_B,\mu_I)$ phase diagram of Ref.\ \cite{He2005}.  The
two lower phase diagrams in Fig.\ \ref{fig:diagrams} (these phase
diagrams were studied in Ref.\ \cite{Warringa2005} with $\mu_u
\leftrightarrow \mu_d$) are of relevance for studying situations which
are in weak equilibrium ($\mu_d = \mu_s$), but not necessarily
electrically neutral. In that case there are situations in which a
charged kaon condensate (j) is competing against a color
superconducting phase (s). From the diagrams it can be seen that the
color superconducting phases are separated from the phase in which the
charged kaon condenses for any temperature and all values of the
chemical potentials. The same conclusion holds for a pion condensate
\cite{Warringa2005}.

\section{Summary and Conclusions}
We presented phase diagrams of the three-flavor NJL model as a
function of flavor chemical potentials and temperature. We found that
the phases with pseudoscalar condensation are separated from the color
superconducting phases by a first order transition for any
temperature. Although many regions in the phase diagrams were or will
never be realized in nature, these diagrams may be relevant for
comparison with possible future lattice data. This work could be
extended with the inclusion of the electric and color neutrality
constraints, the discussion of gapless phases, and the investigation
of the effect of 't Hooft's instanton-induced interaction on the
phases with pseudoscalar condensation.

\section*{Acknowledgments}
This work has been carried out in collaboration with J.O. Andersen and
D. Boer. I would like to thank the organizers of the XQCD'05 workshop
for the stimulating meeting and the opportunity to present this work.


\begin{thebibliography}{99}
\bibitem{Alford2002}
M.~Alford and K.~Rajagopal,
JHEP {\bf 0206}, 031 (2002).

\bibitem{Ruster2005}
S.~B.~Ruster, V.~Werth, M.~Buballa, I.~A.~Shovkovy and D.~H.~Rischke,
%
Phys.\ Rev.\ D {\bf 72}, 034004 (2005).

\bibitem{Blaschke2005}
D.~Blaschke, S.~Fredriksson, H.~Grigorian, A.~M.~{\"O}ztas and F.~Sandin,
%
Phys.\ Rev.\ D {\bf 72}, 065020 (2005).

\bibitem{Son2001}
D.~T.~Son and M.~A.~Stephanov,
Phys.\ Rev.\ Lett.\  {\bf 86}, 592 (2001).

\bibitem{Kogut2002}
J.~B.~Kogut and D.~K.~Sinclair,
Phys.\ Rev.\ D {\bf 66}, 034505 (2002).

\bibitem{Warringa2005}
H.~J.~Warringa, D.~Boer and J.~O.~Andersen,
Phys.\ Rev.\ D {\bf 72}, 014015 (2005).


\bibitem{Toublan2003}
D.~Toublan and J.~B.~Kogut,
%
Phys.\ Lett.\ B {\bf 564}, 212 (2003).

\bibitem{Barducci2004}
A.~Barducci, R.~Casalbuoni, G.~Pettini and L.~Ravagli,
Phys.\ Rev.\ D {\bf 69}, 096004 (2004).

\bibitem{Barducci2005}
A.~Barducci, R.~Casalbuoni, G.~Pettini and L.~Ravagli,
Phys.\ Rev.\ D {\bf 71}, 016011 (2005).

\bibitem{Gastineau2002}
F.~Gastineau, R.~Nebauer and J.~Aichelin,
Phys.\ Rev.\ C {\bf 65}, 045204 (2002).

\bibitem{Neumann2003}
F.~Neumann, M.~Buballa and M.~Oertel,
Nucl.\ Phys.\ A {\bf 714}, 481 (2003).

\bibitem{Lawley2005}
S.~Lawley, W.~Bentz and A.~W.~Thomas,
arXiv:nucl-th/0504020.

\bibitem{Buballa2005a}
M.~Buballa,
Phys.\ Rept.\  {\bf 407}, 205 (2005).

\bibitem{Kogut2001}
J.~B.~Kogut and D.~Toublan,
Phys.\ Rev.\ D {\bf 64}, 034007 (2001).

\bibitem{Alford1999}
M.~G.~Alford, K.~Rajagopal and F.~Wilczek,
Nucl.\ Phys.\ B {\bf 537}, 443 (1999).

\bibitem{Steiner2002}
A.~W.~Steiner, S.~Reddy and M.~Prakash,
Phys.\ Rev.\ D {\bf 66}, 094007 (2002).

\bibitem{Buballa2005c}
M.~Buballa and I.~A.~Shovkovy,
Phys.\ Rev.\ D {\bf 72} 097501 (2005).

\bibitem{Bedaque2002}
P.~F.~Bedaque,
Nucl.\ Phys.\ A {\bf 697}, 569 (2002).

\bibitem{He2005}
L.~y.~He, M.~Jin and P.~f.~Zhuang,
Phys.\ Rev.\ D {\bf 71}, 116001 (2005).

\end{thebibliography}
\end{document}